\documentclass[a4paper]{jpconf}
\usepackage{graphicx}

\usepackage{graphicx} 
\graphicspath{{figures/}} 
\usepackage{booktabs} 
\usepackage[labelfont=bf]{caption} 
\usepackage{amsfonts, amsmath, amsthm, amssymb} 
\usepackage{wrapfig} 
\usepackage[utf8]{inputenc}
\usepackage{textcomp}

\usepackage{feynmp-auto}
\usepackage{enumerate}
\usepackage{textcomp}
\usepackage{longtable}
\usepackage{pdfpages}
\usepackage{float}
\usepackage{fancyhdr} 
\usepackage[font=small,labelfont=bf]{caption}

\newcommand{\maN}{\ensuremath{\mathcal{N}}}
\newcommand{\maT}{\ensuremath{\mathbb{T}}}
\newcommand{\Z}[1]{\ensuremath{\mathbb{Z}_{#1}}} 
\newcommand{\E}[1]{\ensuremath{\mathrm{E}_{#1}}} 
\newcommand{\x}[0]{\ensuremath{\times}}
\newcommand{\I}{\mathrm{i}}
\def\ra{\overline}
\def\fr{\frac}
\def\zc{\bar{z}}
\def\ket{\rangle}

\begin{document}
\title{Kähler potential of heterotic orbifolds with multiple Kähler moduli}

\author{Yessenia Olgu\'in-Trejo, Sa\'ul Ramos-S\'anchez}

\address{Instituto de F\'isica, Universidad Nacional Aut\'onoma de M\'exico,
POB 20-364, Cd.Mx. 01000, M\'exico}

\ead{yess.olt@ciencias.unam.mx, ramos@fisica.unam.mx}

\begin{abstract}

Aiming at improving our knowledge of the low-energy limit of heterotic orbifold compactifications,
we determine at lowest order the Kähler potential of matter fields in the case where more than three 
bulk Kähler moduli appear. Interestingly, bulk matter fields couple to more than one Kähler modulus,
a subtle difference with models with only three Kähler moduli that may provide a tool to address
the question of moduli stabilization in these models.

\end{abstract}


\section{Introduction}

String theory is possibly the prime candidate for an UV-complete and unified description
of particle physics and cosmology. In order to make contact with our observable 4D universe, 
expecting $\maN=1$ supersymmetry at some scale, the six extra dimensions predicted by the 
theory must be compactified on Calabi-Yau manifolds or orbifolds. Below the compactification 
scale, the features of the resulting SUGRA theory are entirely determined by 
the geometry and topology of the compactification space. Thus, one expects 
to identify a compactification space that yields the standard models of particle physics and
cosmology.

Toroidal orbifold compactifications of the \E8\x\E8 heterotic string\footnote{See e.g. 
\cite{Dixon:1985jw,Bailin:1999nk,RamosSanchez:2008tn,Vaudrevange:2008sm} for details on heterotic orbifold 
compactifications in the bosonic formulation.} have shown to lead to large sets
of models that, beyond reproducing properties of the MSSM, provide plausible solutions to 
problems of particle physics~\cite{Kobayashi:2004ya, Buchmuller:2005jr, Kim:2007mt, Nilles:2008gq, Lebedev:2008un, Nilles:2014owa}.
However, to study the cosmological consequences of this scheme one has still to overcome the hurdles posed by
moduli stabilization. Bulk moduli are complex scalar fields parametrizing the size and shape of the
compact space; they have a flat potential at tree level in the effective theory of these models,
leading thus to undesirable properties, such as unobserved fifth forces. Efforts to solve this issue
have been made~\cite{Parameswaran:2010ec, Dundee:2010sb}, revealing that it is still unclear 
whether orbifold compactifications can lead to stable 4D vacua. 

Addressing moduli stabilization requires knowing the details of the associated effective field theory. In
$\maN=1$ preserving heterotic orbifolds, one then needs to determine the superpotential, the gauge kinetic functions
and the Kähler potential. The former two have been computed\footnote{Yet, despite the great progress on 
coupling selection rules~\cite{Kobayashi:2011cw,Bizet:2013gf,Nilles:2013lda,Bizet:2013wha}, there remain 
some details to clarify.} even considering some non-perturbative
contributions~\cite{Burwick:1990tu,Erler:1992gt,Choi:2007nb,Dixon:1990pc,Mayr:1993mq}.
The Kähler potential is less known because, among other things, it is not protected by non-renormalization theorems~\cite{Dine:1986vd},
although it has been explicitly obtained at leading order in some 
cases~\cite{Bailin:1991ts, Kaplunovsky:1995jw}. In this contribution, after briefly reviewing the origin
of bulk moduli in heterotic orbifolds, we reconstruct the Kähler potential 
for one previously unexplored case: when more than three bulk Kähler moduli arise.
To achieve this goal, we present the computation of the relevant scattering amplitudes in the underlying CFT of the heterotic string 
and then compare them to the SUGRA results, which leads to the Kähler metric and thus the Kähler potential.

\section{Untwisted moduli in orbifold compactifications}

{\small
\begin{table}[!b!]
\caption{\label{tab:T1} Twist vectors and Hodge numbers of \Z{n} orbifolds}
\begin{center}
\begin{tabular}{llcccllcc}
\br
\Z{n}    & twist vector         & $h^{1,1}$ & $h^{1,2}$ & \phantom{......}& \Z{n}     & twist vector         & $h^{1,1}$ & $h^{1,2}$\\
\mr
\Z3      & $\fr{1}{3}(1,1,-2)$  & 9         &  0        &                 & \Z8-I     & $\fr{1}{8}(1,2,-3)$  & 3         &  0  \\
\Z4      & $\fr{1}{4}(1,1,-2)$  & 5         &  1        &                 & \Z8-II    & $\fr{1}{8}(1,3,-4)$  & 3         &  1  \\
\Z6-I    & $\fr{1}{6}(1,1,-2)$  & 5         &  0        &                 & \Z{12}-I  & $\fr{1}{12}(1,4,-5)$ & 3         &  0  \\
\Z6-II   & $\fr{1}{6}(1,2,-3)$  & 3         &  1        &                 & \Z{12}-II & $\fr{1}{12}(1,5,-6)$ & 3         &  1  \\
\Z7      & $\fr{1}{7}(1,2,-3)$  & 3         &  0        &                 &  & & & \\
\br
\end{tabular}
\end{center}
\end{table}
}

We focus here on the simplest Abelian toroidal orbifolds, defined by the quotient $\maT^6/\Z{n}$, with
\begin{equation}
  \Z{n} =\{\theta=\vartheta^k\;\;|\;0\leq k< n \},
\end{equation}
where $n$ is restricted by the crystallography of $\maT^6$ to be one of the choices presented in table~\ref{tab:T1}
and the \Z{n} generator $\vartheta$ acts on the $\maT^6$ complex coordinates, $z_i$, $i=1,2,3$, as
\begin{equation}
  \vartheta:\;z_i\ \to\ z_ie^{2\pi\I v_i},\quad 0\leq|v_i|<1,\ i=1,2,3\,,
\end{equation}
where the twist vector $\boldsymbol{v}=(v_1,v_2,v_3)$ is subject to the $\maN=1$ condition $\pm v_1\pm v_2\pm v_3=0$.
In table~\ref{tab:T1} we provide a choice of the twist vectors for all admissible \Z{n} orbifolds.

Invariance of the toroidal metric under the orbifold action leads to the {\it bulk} or {\it untwisted moduli}.
The number of bulk moduli is given by
the Hodge numbers $h^{1,1}$ and $h^{1,2}$, counting respectively Kähler moduli $T$ and complex structure
moduli $U$. Table~\ref{tab:T1} displays the number of modular fields of each kind. Notice particularly 
that, even though most of these orbifolds have three Kähler moduli $T_i$, each associated with one of the coordinates $z_i$,
there can appear up to nine of these moduli, labeled as $T_{ij}$, $i,j=1,2,3$. 

Besides bulk moduli, there appear matter states $\Phi_\alpha$ of two species: {\it bulk matter states}, free to move
everywhere in 10D, and {\it twisted} or {\it localized matter states} that are linked to the singularities of the orbifold, yet
free in the uncompactified space. Matter states build the matter content of the emerging field theory and can reproduce
the properties of MSSM superfields.

In orbifold compactifications without Wilson lines, the effective field theory is invariant
under $SL(2,\mathbb{Z})$ modular transformations of each of the bulk moduli. The resulting target-space modular group
$[SL(2,\mathbb{Z})]_T^{h^{1,1}}\x[SL(2,\mathbb{Z})]_U^{h^{1,2}}$ is broken down to some congruent subgroup
in semi-realistic models due to the inclusion of Wilson lines~\cite{Bailin:1993ri,Love:1996sk}. Bulk and twisted matter states $\Phi_\alpha$ 
transform non-trivially under the modular group~\cite{Ferrara:1989qb,Ibanez:1992hc}, according to their modular
properties encoded in the so-called modular weights $n_\alpha$ and $\ell_\alpha$, for $T$ and $U$ symmetries respectively, 
which are vectors of fractional charges that 
can be determined from the computation of the Kähler potential contributions involving both bulk moduli and matter fields.

\section{Kähler potentials for orbifold compactifications}

The Kähler potential when {\it three} bulk Kähler moduli, none or one complex structure modulus (see table~\ref{tab:T1}) 
and some matter fields are present, is given by~\cite{Ibanez:1992hc} 
\begin{equation}
  \label{eq:1}
  K = -\sum_{i=1}^{3}\ln ({T_{i}+\ra{T}_{i}})  -\delta^{h^{1,2}}_1\ln(U+\ra{U})
      +\sum_\alpha\prod_{i=1}^{3} (U+\ra{U})^{-\ell_\alpha} (T_{i}+\ra{T}_{i})^{-n_\alpha^i} \Phi_\alpha\ra{\Phi}_{\ra\alpha}\,.
\end{equation}  
The modular weights of an untwisted matter field $\Phi_\alpha$ are $n_\alpha^i = -q_\alpha^i$ and $\ell_\alpha = -q_\alpha^3$, 
with $q_\alpha$ denoting the Neveu-Schwarz right-moving momentum of the field (with entries $0$ or $-1$); 
and for twisted states $\Phi_\alpha$ of the $\theta$-sector they are given by
\begin{equation}
  \label{eq:modularweights}
  \begin{array}{llcll}
  n_\alpha^i = 1-\eta_\alpha^i -\Delta\tilde N_\alpha^i\,,& \ell_\alpha = 1-\eta_\alpha^3 +\Delta\tilde N_\alpha^3 &\qquad&\text{for } \eta_\alpha^i\neq0\,, &\\
  n_\alpha^i = \eta_\alpha^i\,,                           & \ell_\alpha = \eta_\alpha^3                           &       &\text{for } \eta_\alpha^i = 0,    & i=1,2,3\,, 
  \end{array}
\end{equation}
where $\Delta\tilde N_\alpha^i$ denotes the number of holomorphic minus antiholomorphic left-moving oscillator excitations 
in the $z^i$ complex plane generating the state, and $\eta^i = k v_i\mod 1$, such that $0\leq\eta^i<1$, in the 
$\theta=\vartheta^k$ twisted sector. $\ell_\alpha=0$  $\forall\alpha$ if $h^{1,2}=0$.

For models with more than three Kähler moduli, we expect $K$ to have a similar form, but the slight differences 
may be relevant for moduli stabilization. In general, denoting any bulk modulus as $M$, the Kähler potential of interest
can be written as an expansion of the matter fields $\Phi$ as 
\begin{equation}
\label{eq:0}
  K(M,\ra{M},\Phi,\ra{\Phi})=\mathcal{K}(M,\ra{M})+\sum_{\alpha,\beta}K,_{\alpha\bar\beta}\Phi^\alpha\ra\Phi{}^{\bar\beta}+...,
\end{equation}
where $\mathcal{K}$ is a function that has been determined through dimensional reduction of 
10D SUGRA theories~\cite{Ferrara:1986qn} as it only depends on the effective behavior of the moduli. 
On the other hand, $K,_{\alpha\bar\beta}$ is an unknown function of the bulk moduli that
describes the coupling between moduli and matter fields, that carry full information from the original string theory.

In string models, this function can be obtained at first order from four-point scattering 
amplitudes~\cite{Bailin:1991ts, Kaplunovsky:1995jw}, which are determined by the correlation functions
between the asymptotic states $|\Phi\ket$ and $|\Phi'\ket$ corresponding to the matter fields $\Phi$ and $\Phi'$,
affected by the interaction with the vertex operators $V_M$ and $V_{M'}$ associated with the moduli $M$ and $M'$.
Schematically, the amplitude to compute reads 
$A(M,\Phi,(\Phi')^\dagger,(M')^\dagger) := \frac1{4\pi}\int dz^2\langle \Phi'|V_{\ra{M}'}V_M|\Phi\ket$.

The vertex operator of a Kähler modulus $M_{a\bar{e}}$ with components in the $z^a$ and $\bar z^{\bar e}$ complex directions, 
and in the 0-ghost picture, takes the form (we set $\alpha'=1/2$)~\cite{Hamidi:1986vh} 
\begin{equation}
\label{eq:4}
  V_{0}(M_{a\bar{e}},k,z,\zc)=\left(\partial_{z}X^{a}_{R}+\frac{1}{2}\Psi^{a}_{R}k\cdot\Psi_{R}\right) 
                                   e^{ik\cdot X_{R}} \partial_{\zc}X_{L}^{\bar{e}}e^{ik\cdot X_{L}},
\end{equation}
where $X$ and $\Psi$ are the bosonic and fermionic worldsheet fields and $k$ is the momentum of $M$, Latin indices are $1,2,3$.
On the other hand, the Neveu-Schwarz components of the matter states with gauge momentum $p_\alpha$
are given by~\cite{Bailin:1992he}
\begin{equation}
\label{eq:3}
\begin{array}{rlclcl}
  |\Phi_\alpha\ket &= b^{\alpha_i}_{-1/2} |0\ket_R \otimes|p_\alpha\ket_L\,,                &\quad& \alpha_i=1,2,3 &\quad &\text{untwisted sector},  \\
  |\Phi_\alpha\ket &= \Omega_\alpha|0\ket_R \otimes \widetilde\Omega_\alpha|p_\alpha\ket_L  &     &                &      & \text{twisted sector},
\end{array}
\end{equation}
where $\Omega_\alpha$ and $\widetilde\Omega_\alpha$ are products of right and left-handed oscillators, respectively, in possibly
various complex directions. Consequently, in contrast to untwisted states, a twisted state $|\Phi_\alpha\ket$ can carry more 
than one index $i=1,2,3$ and many oscillators in various directions.

Using the canonical formalism in CFT, we can now compute the correlation function associated with the scattering 
of two untwisted matter fields and two Kähler moduli. After a lengthy computation, the result reads
\begin{equation}
\label{eq:6}
\text{\small
\mbox{$
  A(M_{a\bar{e}},\Phi_{\gamma},(\Phi_{\beta})^{\dagger},(M_{b\bar{f}})^{\dagger})=
  \frac{1}{32}\delta_{\bar{e}\bar{f}}\left(\delta_{\gamma\beta}\delta_{ab}\frac{us}{t}+
  s\delta_{\gamma_i a} \delta_{\beta_i b}\delta_{\gamma\beta}\right)
  \frac{\Gamma(1-u/8)\Gamma(1-s/8)\Gamma(1-t/8)}{\Gamma(1+u/8)\Gamma(2+t/8)\Gamma(1+s/8)},
  $}
  }
\end{equation}
where $s,u,t$ are the kinematic Mandelstam variables of the process. We are interested in this result at
low energies, i.e. when $|s|,|u|,|t|\ll1$. In this case,~\eqref{eq:6} takes the form
\begin{equation}
\label{eq:7}
  A(M_{a\bar{e}},\Phi_{\gamma},(\Phi_{\beta})^{\dagger},(M_{b\bar{f}})^{\dagger}) \approx
  \frac{1}{32}\delta_{\bar{e}\bar{f}}\left(\delta_{\gamma\beta}\delta_{ab}\frac{us}{t}+
  s\delta_{\gamma_i a}\delta_{\beta_i b}\delta_{\gamma\beta}\right).
\end{equation}

Now, in SUGRA one can also determine the analogous amplitude in terms of the fields instead of strings 
and then express it in the small-moduli limit ($M\ll1$), by generalizing the methods developed in~\cite{Dixon:1989fj}. 
$A$ is given thus by 
\begin{equation}
\label{eq:8}
  A(M_{a\bar{e}},\Phi_{\gamma},(\Phi_{\beta})^{\dagger},(M_{b\bar{f}})^{\dagger}) \approx
     \frac{1}{32}\left(\frac{us}{t}\delta_{ab}\delta_{\bar{e}\bar{f}}\delta_{\gamma\beta}+sG_{\gamma\bar{\beta};a\bar{e},b\bar{f}}(0,0)\right)\,,
\end{equation}
where $G_{\gamma\bar{\beta};a\bar{e},b\bar{f}}(0,0)$ is the derivative of the Kähler metric with
respect to the moduli $M_{a\bar e}$ and $(M_{b\bar f})^\dagger$ in the small-moduli limit.
Contrasting~\eqref{eq:7} and~\eqref{eq:8}, we can readily conclude that the Kähler metric is
\begin{equation}
\label{eq:result1}
   G_{\gamma\bar{\beta}}\approx \delta_{\gamma\beta}
                         +\delta_{\gamma\beta}\delta_{\bar{e}\bar{f}}\delta_{\gamma_i a} \delta_{\beta_i b}M_{a\bar{e}}(M_{b\bar{f}})^{\dagger}\,,
\end{equation}
which is, by definition, the function $K,_{\gamma\bar\beta}$ in~\eqref{eq:0}. 

From~\eqref{eq:result1}, we find that, if the in and out states coincide, i.e. $\gamma=\beta$, 
then $\gamma_i=\beta_i$ and, after the holomorphic field redefinition $M_{i\bar\jmath} \to 1-T_{ij} /1+T_{ij}$,
a Kähler transformation and taking the small-moduli limit, the relevant contribution to the Kähler potential reads 
\begin{equation}
\label{eq:untwistedK}
  K \supset \sum_{e}\left(1 + M_{\beta_i \bar e} (M_{\beta_i \bar e})^\dagger\right)|\Phi_\beta|^2 
    \quad\longrightarrow\quad  \prod_{e} (T_{\beta_i e} + \ra T_{\beta_i e})^{-1} |\Phi_\beta|^2\,.
\end{equation}
Note that, in contrast to dimensional reductions of 10D SUGRA~\cite{Ferrara:1986qn},
orbifold compactifications do not yield terms proportional to $\Phi_\beta\ra\Phi_{\bar\gamma}$, with $\beta\neq\gamma$.
The modular weights of $\Phi_\beta$ in the untwisted sector are directly read off~\eqref{eq:untwistedK}. They build 
now the matrix $n_\beta^{ij}=\delta_{\beta_i}^i$ $\forall j$, where the values that $i,j$ 
take depend on the Kähler moduli present in the model. In the \Z4 and \Z6-I orbifolds, only $T_{ij}$ with $i,j=1,2$ and $T_{33}$ exist.

The computation for the twisted sectors is analogous. There are several sources of differences though. First, the expressions
for $X_R^a$, $\Psi_R^a$ and $X_L^{\bar e}$ in the vertex operator~\eqref{eq:4} of a modulus $M_{a\bar e}$ depend on two
components of the twist, $\eta^a$ and $\eta^{\bar e}=\eta^e$. Secondly, the asymptotic states $|\Phi_\alpha\ket$ carry 
non-trivial oscillators (with frequencies that depend on the twist) in various directions. Massless states exhibit
holomorphic and antiholomorphic oscillators in $X_L$, whose difference in the compact direction $z^i$ is
denoted $\Delta\tilde{N}^i_\alpha$; right-moving bosonic oscillators in $\Psi_R$ of the type $b^i_{\nu}$,
for some frequencies $\nu=\nu(\eta^i)$, affect the correlators; and there are no oscillators from $X_R$ in massless states.
These features leave their
trace in the expressions of the correlators. Setting first $\Delta\tilde{N}^i_\alpha=0$, after a lengthy computation, we arrive at
\begin{equation}
\label{eq:10}
\text{\small\mbox{
 $A(M_{a\bar e},\Phi_{\beta},\Phi_{\gamma}^{\dagger},(M_{b\bar f})^{\dagger})=
    \delta_{\gamma\beta}\delta_{ab}\delta_{\bar e\bar f}\frac{s}{32t}(u+t(1-\eta_\beta^{\bar e}))
    \frac{\Gamma(1-\eta_\beta^{\bar e}-u/8)\Gamma(\eta_\beta^a-s/8)\Gamma(1-t/8)}{\Gamma(\eta_\beta^{\bar e}+u/8)\Gamma(2+t/8)\Gamma(1+s/8-\eta_\beta^{\bar e})}.$
}}
\end{equation}
Including oscillators and taking its low-energy limit (using that $\eta^a_\beta = \eta^{\bar e}_\beta$ for non-diagonal moduli)
yields
\begin{equation}
\label{eq:11}
A(M_{a\bar{e}},\Phi_{\beta},\Phi_{\gamma}^{\dagger},(M_{b\bar{f}})^{\dagger}) \approx
\delta_{\gamma\beta}\delta_{ab}\delta_{\bar{e}\bar{f}}\frac{s}{32t}\left(u+t (1 - \eta_\beta^{a} + \Delta\tilde{N}_\beta^{a})\right).
\end{equation}

We must now compare this result with the SUGRA analogue, to identify the Kähler metric, which for twisted states turns out to be
\begin{equation}
\label{eq:result2}
   G_{\gamma\bar{\beta}}\approx \delta_{\gamma\beta}
                         +\delta_{\gamma\beta}\delta_{ab}\delta_{\bar{e}\bar{f}} (1 - \eta_\beta^{a} + \Delta\tilde{N}_\beta^{a}) M_{a\bar{e}}(M_{b\bar{f}})^{\dagger}\,.
\end{equation}
As in the untwisted case, by performing the proper transformations and considering  
the small-moduli limit, we finally obtain
\begin{equation}
\label{eq:twistedK}
  K \supset \sum_{a,e}\left(1 + n_\beta^{ae} M_{a \bar e} (M_{a \bar e})^\dagger\right)|\Phi_\beta|^2 
    \quad\longrightarrow\quad  \prod_{a,e} (T_{a e} + \ra T_{a e})^{-n_\beta^{ae}} |\Phi_\beta|^2\,,
\end{equation}
where we have identified the twisted modular weights as $n_\beta^{ae}=1 - \eta_\beta^{a} + \Delta\tilde{N}_\beta^{a}$ $\forall e$.
We realize that~\eqref{eq:untwistedK} and~\eqref{eq:twistedK} are very similar, but they have two differences. The first one
is the expression for the modular weights, given under these equations. The second one is that the Kähler contribution
of the twisted fields prescribes a kinetic interaction between these fields and all Kähler moduli, whereas the
untwisted fields interact only with a subset of the moduli. The latter is also an interesting difference with respect
to compactifications with only three bulk moduli: each bulk matter field interacts only with one Kähler modulus in those models.

From our discussion, we find that the Kähler potential at leading order in heterotic orbifold compactifications endowed 
with more than the three diagonal Kähler moduli takes the form
\begin{equation}
  \label{eq:FinalResult}
  K = -\ln \det({T_{ij}+\ra{T}_{ij}}) -\delta^{h^{1,2}}_1\ln(U+\ra{U})
      +\sum_\beta\prod_{i,j} (U+\ra{U})^{-\ell_\beta} (T_{i j}+\ra{T}_{i j})^{-n_\beta^{i j}} \Phi_\beta\ra{\Phi}_{\bar\beta}\,,
\end{equation}
where the Kähler modular weights are defined below~\eqref{eq:untwistedK} and~\eqref{eq:twistedK}, and the 
complex structure modular weights are given in~\eqref{eq:modularweights}. The first contribution in $K$ 
is a result of dimensional reduction in SUGRA~\cite{Ferrara:1986qn}, which applies here because bulk moduli of
heterotic orbifolds behave just as in a SUGRA compactification.

\section{Discussion and forthcoming research}

This proceedings contribution has been devoted to ameliorating our understanding of the SUGRA limit
of supersymmetric heterotic orbifold compactifications. We have
reconstructed the Kähler potential at lowest order in the case where bulk and twisted matter fields interact with
more than the three usual bulk Kähler moduli. To do so, we computed the associated four-point scattering amplitudes within the
formalism of the underlying CFT of the heterotic string and compared them with the known analogous result in SUGRA.

Our findings are summarized in eq.~\eqref{eq:FinalResult}. In the presence of multiple bulk Kähler moduli in orbifolds,
the structure of the couplings between matter fields and bulk matter fields becomes richer than in models with
only diagonal Kähler moduli. Especially, each bulk matter field is coupled to more than one Kähler modulus.
This feature may become instrumental to address the longstanding issue of moduli stabilization,
which might only find a solution in a model-dependent basis,
in promising \Z{4}, \Z{6}-I heterotic orbifolds~\cite{Nilles:2014owa} with more than three moduli. 
From there, one can aim at gaining some insight into the string cosmology of these constructions. 

Our result also reveals further details about the modular properties of the effective fields in heterotic orbifolds. 
Provided that the modular weights can be regarded as charges of the target space modular symmetries, on may 
ask oneself if they can play a role in some 4D particle phenomenology, as it has been sometimes suggested (see e.g.~\cite{Feruglio:2017spp}).

Our results are only the starting point of a long quest, whose progress shall be reported elsewhere. We should be able to compute further corrections
of the Kähler potential and thereby try to identify some symmetries governing its structure. The final goal would be to
configure the final shape of the effective SUGRA theory in these simple compactifications. 

\section*{Acknowledgments}
We thank P.K.S Vaudrevange and M. Ratz for interesting discussions on this subject.
This work was partly supported by DGAPA-PAPIIT grant IN100217 and CONACyT grant F-252167.
S.~R-S. would like to thank the ICTP for the kind hospitality and support received
through its Junior Associateship Scheme during the realization of this work.

\section*{References}

\begin{thebibliography}{10}
\expandafter\ifx\csname url\endcsname\relax
  \def\url#1{{\tt #1}}\fi
\expandafter\ifx\csname urlprefix\endcsname\relax\def\urlprefix{URL }\fi
\providecommand{\eprint}[2][]{\url{#2}}

\bibitem{Dixon:1985jw}
Dixon L~J, Harvey J~A, Vafa C and Witten E 1985 {\em Nucl. Phys.\/} {\bf B261}
  678--686

\bibitem{Bailin:1999nk}
Bailin D and Love A 1999 {\em Phys. Rept.\/} {\bf 315} 285--408

\bibitem{RamosSanchez:2008tn}
Ramos-S{\'a}nchez S 2009 {\em Fortsch. Phys.\/} {\bf 10} 907--1036
  (\textit{Preprint} \eprint{0812.3560})

\bibitem{Vaudrevange:2008sm}
Vaudrevange P~K~S 2008 Ph.D. thesis Bonn U. (\textit{Preprint}
  \eprint{0812.3503})

\bibitem{Kobayashi:2004ya}
Kobayashi T, Raby S and Zhang R~J 2005 {\em Nucl. Phys.\/} {\bf B704} 3--55
  (\textit{Preprint} \eprint{hep-ph/0409098})

\bibitem{Buchmuller:2005jr}
Buchm{\"u}ller W, Hamaguchi K, Lebedev O and Ratz M 2006 {\em Phys. Rev.
  Lett.\/} {\bf 96} 121602 (\textit{Preprint} \eprint{hep-ph/0511035})

\bibitem{Kim:2007mt}
Kim J~E, Kim J~H and Kyae B 2007 {\em JHEP\/} {\bf 06} 034 (\textit{Preprint}
  \eprint{hep-ph/0702278})

\bibitem{Nilles:2008gq}
Nilles H~P, Ramos-S{\'a}nchez S, Ratz M and Vaudrevange P~K~S 2009 {\em Eur.
  Phys. J.\/} {\bf C59} 249--267 (\textit{Preprint} \eprint{0806.3905})

\bibitem{Lebedev:2008un}
Lebedev O, Nilles H~P, Ramos-S{\'a}nchez S, Ratz M and Vaudrevange P~K~S 2008
  {\em Phys. Lett.\/} {\bf B668} 331--335 (\textit{Preprint}
  \eprint{0807.4384})

\bibitem{Nilles:2014owa}
Nilles H~P and Vaudrevange P~K~S 2015 {\em Mod. Phys. Lett.\/} {\bf A30}
  1530008 (\textit{Preprint} \eprint{1403.1597})

\bibitem{Parameswaran:2010ec}
Parameswaran S~L, Ramos-S{\'a}nchez S and Zavala I 2011 {\em JHEP\/} {\bf 01}
  071 (\textit{Preprint} \eprint{1009.3931})

\bibitem{Dundee:2010sb}
Dundee B, Raby S and Westphal A 2010 {\em Phys. Rev.\/} {\bf D82} 126002
  (\textit{Preprint} \eprint{1002.1081})

\bibitem{Kobayashi:2011cw}
Kobayashi T, Parameswaran S~L, Ramos-S{\'a}nchez S and Zavala I 2012 {\em
  JHEP\/} {\bf 05} 008 [Erratum: JHEP12,049(2012)] (\textit{Preprint}
  \eprint{1107.2137})

\bibitem{Bizet:2013gf}
Cabo~Bizet N~G, Kobayashi T, Mayorga~Pe{\~n}a D~K, Parameswaran S~L, Schmitz M
  and Zavala I 2013 {\em JHEP\/} {\bf 05} 076 (\textit{Preprint}
  \eprint{1301.2322})

\bibitem{Nilles:2013lda}
Nilles H~P, Ramos-Sánchez S, Ratz M and Vaudrevange P~K~S 2013 {\em Phys.
  Lett.\/} {\bf B726} 876--881 (\textit{Preprint} \eprint{1308.3435})

\bibitem{Bizet:2013wha}
Cabo~Bizet N~G, Kobayashi T, Mayorga~Pe{\~n}a D~K, Parameswaran S~L, Schmitz M
  and Zavala I 2014 {\em JHEP\/} {\bf 02} 098 (\textit{Preprint}
  \eprint{1308.5669})

\bibitem{Burwick:1990tu}
Burwick T~T, Kaiser R~K and M{\"u}ller H~F 1991 {\em Nucl. Phys.\/} {\bf B355}
  689--711

\bibitem{Erler:1992gt}
Erler J, Jungnickel D, Spalinski M and Stieberger S 1993 {\em Nucl. Phys.\/}
  {\bf B397} 379--416 (\textit{Preprint} \eprint{hep-th/9207049})

\bibitem{Choi:2007nb}
Choi K~S and Kobayashi T 2008 {\em Nucl. Phys.\/} {\bf B797} 295--321
  (\textit{Preprint} \eprint{0711.4894})

\bibitem{Dixon:1990pc}
Dixon L~J, Kaplunovsky V and Louis J 1991 {\em Nucl. Phys.\/} {\bf B355}
  649--688

\bibitem{Mayr:1993mq}
Mayr P and Stieberger S 1993 {\em Nucl. Phys.\/} {\bf B407} 725--748
  (\textit{Preprint} \eprint{hep-th/9303017})

\bibitem{Dine:1986vd}
Dine M and Seiberg N 1986 {\em Phys. Rev. Lett.\/} {\bf 57} 2625

\bibitem{Bailin:1991ts}
Bailin D, Gandhi S~K and Love A 1992 {\em Phys. Lett.\/} {\bf B275} 55--62

\bibitem{Kaplunovsky:1995jw}
Kaplunovsky V and Louis J 1995 {\em Nucl. Phys.\/} {\bf B444} 191--244
  (\textit{Preprint} \eprint{hep-th/9502077})

\bibitem{Bailin:1993ri}
Bailin D, Love A, Sabra W~A and Thomas S 1994 {\em Mod. Phys. Lett.\/} {\bf A9}
  1229--1238 (\textit{Preprint} \eprint{hep-th/9312122})

\bibitem{Love:1996sk}
Love A and Todd S 1996 {\em Nucl. Phys.\/} {\bf B481} 253--288
  (\textit{Preprint} \eprint{hep-th/9606161})

\bibitem{Ferrara:1989qb}
Ferrara S, L{\"u}st D and Theisen S 1989 {\em Phys. Lett.\/} {\bf B233}
  147--152

\bibitem{Ibanez:1992hc}
Ib{\'a\~n}ez L~E and L{\"u}st D 1992 {\em Nucl. Phys.\/} {\bf B382} 305--361
  (\textit{Preprint} \eprint{hep-th/9202046})

\bibitem{Ferrara:1986qn}
Ferrara S, Kounnas C and Porrati M 1986 {\em Phys. Lett.\/} {\bf B181} 263

\bibitem{Hamidi:1986vh}
Hamidi S and Vafa C 1987 {\em Nucl. Phys.\/} {\bf B279} 465--513

\bibitem{Bailin:1992he}
Bailin D and Love A 1992 {\em Phys. Lett.\/} {\bf B288} 263--268

\bibitem{Dixon:1989fj}
Dixon L~J, Kaplunovsky V and Louis J 1990 {\em Nucl. Phys.\/} {\bf B329} 27--82

\bibitem{Feruglio:2017spp}
Feruglio F 2017  (\textit{Preprint} \eprint{1706.08749})

\end{thebibliography}

\providecommand{\newblock}{}

\end{document}